\documentclass[12pt,letterpaper]{article}
\pdfoutput=1
\usepackage{amssymb}
\usepackage{cite}
\usepackage{amsmath}
\usepackage{amsfonts}
\usepackage{epsfig}
\usepackage{color}
\usepackage{hyperref}
\usepackage{tikz-feynman}
\usepackage{tikz, pgf}
\usetikzlibrary{shapes.misc}
\usetikzlibrary{snakes}
\usetikzlibrary{decorations.pathmorphing}
\usetikzlibrary{shapes}
\usetikzlibrary{fit}

\usepackage{bm}
\outer\def\beginsection#1\par{\medbreak\bigskip
      \message{#1}\leftline{\bf#1}\nobreak\medskip
\vskip-\parskip
      \noindent}

\def\be{\begin{equation}}
\def\ee{\end{equation}}
\def\ba{\begin{eqnarray}}
\usepackage[titles]{tocloft}

\usepackage{titlesec}
\titleformat*{\section}{\large  \bfseries }
\titleformat*{\subsection}{\normalsize  \bfseries }

\newenvironment{fleqnarray*}
  {\@fleqntrue\begin{eqnarray*}}
  {\end{eqnarray*}\@fleqnfalse}
\begin{document}
%%%%%%%%%%%%%%%%%%%%%%%
\begin{titlepage}
\vspace*{1cm}
\vskip 2.5cm
%\vskip 1.0cm
\begin{center}
{\Large \bf
Soft-Theorems for Scalar Particles: The Dilatons Story}
 
\vskip 1.0cm 
{\large Raffaele Marotta} \\[0.7cm] 
%{\it $^a$}\\
%{\it $^b$ }\\[2mm]
{\it  Istituto Nazionale di Fisica Nucleare, Sezione di Napoli, Complesso \\
 Universitario di Monte S. Angelo ed. 6, via Cintia, 80126, Napoli, Italy}\\[2mm]
\end{center}
\vskip 1.0cm 

\begin{abstract}
We summarize  recent results  regarding    single and double soft theorems of two different particles named dilatons, the Nambu-Goldstone boson of the spontaneously broken conformal field theories and the massless scalar particle of the closed string theories. Similarities and differences between the soft theorems of these two particles are discussed as well as their connections with the symmetries of the theories.
\end{abstract}
\end{titlepage}
\section{Introduction}
Symmetries are one of the most powerful guiding principle for constructing consistent mathematical frameworks necessary to  describe quantitatively the fundamental laws of nature. All the theories of the known interactions are based on such a principle and it turns out to be  of great interest to understand how the symmetries, eventually hidden,
are manifest in the behaviour of the physical observables.
It is well known, for example, from the pioneering works by Low \cite{Low1,Low2,Low3,Low4,Low5} for gauge theories and Weinberg \cite{Weinberg1,Weinberg2,Weinberg3} in gravity that gauge symmetries determine the behaviour of the S-matrix 
 in the infrared region where gauge-bosons or gravitons  carry low or soft momentum \cite{gauge1,gauge2,DiVecchia:2016szw}. In this  soft-regime,  scattering amplitudes with a soft graviton or  gauge boson factorize in the product of amplitudes with only hard or finite energy  particles 
 and a soft contribution. 
This behaviour  is   universal at leading order in the low momentum expansion and for tree level amplitudes extends to subleading order  for amplitudes with soft gravitons \cite{DiVecchia:2016szw,1611.07534,uni1,uni2,uni3,uni4,uni5,uni6,uni7,uni8,uni8b,uni9,uni10,uni11,uni12,uni13,uni14,uni15,uni16,uni17,uni18}.  Loop amplitudes in four space-time dimensions, instead, are usually ill-defined because of the presence of infrared divergences. These divergences make even the leading soft factor not universal for Yang-Mills amplitudes \cite{1405.1015} while for gravity and abelian theories their effect consists in adding logarithmic corrections to the subleading soft factors \cite{Weinberg:Photon,log1,log2,log3}. 
The soft regime of amplitudes with gravitons also provides predictions on the gyromagnetic factors of specific massive higher spin states emerging in Kaluza-Klein theories \cite{1911.05099}
that they have been  
 confirmed by explicit amplitude  calculations \cite{2102.13180}.

The deep relation between symmetries and IR-behaviour of the $S$-matrix, have been recently extended to the asymptotic 
 symmetries of  flat space-times introduced by Bondy, Metzner, van de Burg and Sachs to study the gravitational radiation at null infinity 
 (BMS-group) \cite{BMBS1,BMBS2}. The Ward-identities associated 
with the supertranslation generators of the BMS group give the leading soft behaviour of  amplitudes with external gravitons while the subleading contribution originates from the super-rotation charges \cite{Strominger}.

Symmetries determine also the IR behaviour of scattering processes with massless scalar states \cite{Ademollo:1975pf,Shapiro:1975cz,DMM1,DMMN,DMM2,1703.07885,1808.04845,1907.01036} %(having low momenta) 
although,  recently, an alternative geometric proof of such theorems for scalar particles has been proposed in Ref. \cite{geometric}. It is well known, for example, that  when an internal symmetry is spontaneously broken, as in  the non-linear $\sigma$-models (NLSM),  the conservation law of the broken current requires the vanishing of  amplitudes with emission at zero momentum  of a Goldstone boson(Adler's zero)~\cite{Adler:1964um, Adler:1965ga,Weinberg:1966gjf,2009.07817}. 

The situation is different when a space-time symmetry is spontaneously broken as  in the case of the breaking of the conformal group to the Poincar\'e symmetry. In this breaking  only one Nambu-Goldstone boson, in the following named CFT-dilaton,  appears in the spectrum of the theory \cite{0110285}. Amplitudes with a CFT-dilaton and an arbitrary number of massless states are vanishing to leading order in the soft expansion of the dilaton momentum, but for massive hard states, differently from the Adler-zero case, these are divergent at zero dilaton momentum in full analogy with emission of soft-gauge bosons \cite{DMMN}.  To the subleading order instead,  amplitudes with one soft dilaton are determined by the conformal Ward-identity of broken scale invariance \cite{Callan:1970yg,Coleman:1970je,Boels:2015pta} while only recently it has been shown that the subsubleading behaviour of soft dilaton amplitudes is completely fixed from the Ward-identity associated with the breaking of the special conformal generators \cite{DMMN,1705.06175}. These relations hold for any non-anomalous conformal field theory and they have been tested in ${\cal N}=4$ super-Yang Mills perturbatively up to one-loop and non-perturbatively in the contest of gravity/dual on the Coulomb branch and by considering instanton effective actions \cite{1705.06175,Bianchi:2016viy}. 

Symmetries are also manifest in the double soft behaviour of amplitudes computed in NLSM-models \cite{Weinberg:1966gjf,Dashen:1969ez,ArkaniHamed:2008gz,Kampf:2013vha,Low:2015ogb,Du:2015esa}.  Amplitudes with two soft Nambu-Goldstone bosons factorize in amplitudes without the soft particles and a contribution that capture the algebra of the broken generators.
For the dilaton, the double soft behaviour follows from the Ward-identities associated with the breaking of the dilatation and special conformal transformation generators. Furthermore, it turns out to be equal to two consecutive single soft \mbox{expansions \cite{1705.06175}.}

There is another particle named dilaton, it is the massless scalar of the gravitational multiplet of the closed string theories. Amplitudes with this dilaton, referred to in these notes as gravity dilaton, satisfy single and double soft-theorems, as well. These have been determined by computing amplitudes in bosonic, heterotic and superstring theories and expanding them in the relevant infrared region \cite{DiVecchia:2016szw,Ademollo:1975pf, DMM1,DMM2,1808.04845,1907.01036}. 

The same behaviour has also been derived with a different approach based on the property of the string theory to provide a unified expression for amplitudes with external gravitons, dilaton and Kalb-Ramonds states. The request of gauge invariance fixes,  up to subsubleading order in soft expansion, the soft behaviour of the whole amplitude including the dilaton contribution \cite{DiVecchia:2016szw,DiVecchia:2017gfi}. This contribution in the case of a single soft gravity dilaton is conjectured to be universal being the same in all the cases studied and shares many features with those of the CFT-dilatons. This similarity has also led to argue the presence of hidden tree-level conformal symmetry in ordinary  Einstein gravity \cite{1802.05999}.  
However, the resemblance of the two theorems is less evident in the case  of two soft dilatons  because for the gravity dilaton, differently from the CFT-dilaton, the double soft behaviour   cannot be obtained from two consecutive single soft expansions due to the presence of four-point interactions \cite{2005.05877}.

The aim of this article, based on the results obtained in the papers of Ref. \cite{DiVecchia:2016szw,DMM1,DMMN,DMM2, 1705.06175,DiVecchia:2017gfi,2005.05877} is to review  the state of art of the single and double  soft-theorems satisfied by dilaton amplitudes. It will be given  the main  results achieved in the literature providing few details on their derivations but highlighting similarities and differences between the IR-behaviour of the amplitudes with the two dilatons.

The paper is organized as follows. In the Section \ref{Prelude},  are summarized the main features of conformal field theories necessary to introduce  the results considered in the article.  In its subsection are discussed the Ward-identities of the conformal field theory spontaneously broken and their connections with the infrared properties of amplitudes with CFT-dilatons. In Section \ref{String} are introduced the main properties of the string amplitudes used in the subsection to discuss the soft theorems for the gravity-dilatons. In the discussion section are commented  the achieved results.

%%%%%%%%%%%%%%%%%%%%%%%%%%%%%%%%%%%%%%%%%%
\section{CFT-prelude}
\label{Prelude}
A conformal transformation is an invertible map $x\rightarrow x'$ that leave the metric invariant up to a scale factor 
$g_{\mu\nu}(x)\rightarrow \Lambda(x) g_{\mu\nu}(x)$ \footnote{More details on conformal %MDPI: Footnote is not permitted in this journal, so we have moved it into the text, please confirm the whole text.
 field theories and their breaking can be found in  Refs. \cite{DiFrancesco,Higashijima:1994,0110285}.}. The conformal group is locally isomorphic to $SO(2,d)$ with $d$ the space time dimensions and it can be seen as an extension with dilatations ${\cal D}$ and special conformal transformations $\mathcal{K}_\mu$ of the Poincar\`e group.   The generators are:
\begin{eqnarray}
&&\mathcal{P}^\mu   = i \partial^\mu  \hspace*{2.7cm} \mathcal{J}^{\mu \nu} 
= - i (x^\mu \partial^\nu - x^\nu \partial^\mu ) - \mathcal{S}^{\mu \nu}
\nonumber\\
&&\mathcal{D} = i(\Delta + x_\mu \partial^\mu)
\, \hspace*{1.1cm}
\mathcal{K}^\mu 
=i(2 x^\mu x_\nu \partial^\nu - x^2\partial^\mu + 2\Delta x^\mu) + 2x_\nu \mathcal{S}^{\mu \nu}  
\end{eqnarray}
with $\mathcal{P}^\mu$ and $\mathcal{J}^{\mu \nu}$ the generators of the translations and Lorentz transformations respectively and ${\cal S}^{\mu\nu}$ the spin angular momentum operator. $\Delta$ denotes the scaling dimension of the fields involved in the transformation.
The action of the dilation and special transformation on a scalar field $\phi(x)$ with conformal dimension $\Delta$ is:
\begin{eqnarray} 
&&\delta \phi (x ) = [ {\cal D} , \phi (x) ]= i 
 \left( \Delta + 
 x^\mu \partial_\mu 
\right) \phi (x)\nonumber\\
&&\delta_{(\lambda)} \phi (x) 
= \left[\mathcal{K}_{\lambda},\phi(x)\right] 
= i
 \left[(2 x_\lambda x_{\nu} - \eta_{\lambda \nu} x^2 ) \partial^\nu 
+ 2\,\Delta\,x_\lambda \right]\phi(x)\,.
\end{eqnarray}

The %MDPI: can add indentation? please check all conention after euqation, please confirm and revise.
 N\"other currents associated to the scale and  special conformal transformations
\begin{eqnarray}
J^\mu_{\cal D}=x_\nu\,T^{\mu\nu}\qquad (J_{\mathcal{K}})^\mu_{\rho}=(2x_\nu\,x_\rho -\eta_{\rho\nu}\,x^2)
T^{\mu\nu}
\end{eqnarray}
are conserved if the improved energy momentum tensor $T^{\mu\nu}$ is conserved and traceless.

The conformal symmetry is spontaneosly broken  when a scalar field acquires a non zero vacuum expectation value (vev).

The vacuum remains invariant under 
the Poncar\'e group 
but as consequence of the Goldstone's theorem, 
one massless boson, the CFT-dilaton $\xi(x)$, appears in the spectrum parametrizing the oscillations  around the nonconformal vacuum. The dilaton couples linearly to the energy momentum tensor of the broken theory  and through the equations of motion contributes to  its trace:
\begin{eqnarray}
T_\mu^\mu = -v \, \partial^2 \xi \ ,
\end{eqnarray}   
 where $v$  is a dimensionful constant related to the vev of the scalar field. As a consequence, the currents of the broken generators are not conserved being their divergences related to the trace of the energy momentum tensor through the identies:
 \begin{eqnarray}
 \partial_{\mu}\,j_{\cal D}^{\mu}(x) =T_\mu^{~\mu}(x)= -{v}\,\partial^2\,\xi (x)~~;~~\partial_\mu\,j^\mu_{\phantom{\mu}(\lambda)} &= 2\,x_{\lambda}\,T^{\mu}_{\phantom{\mu}\mu}
=-2\,{v}\,x_{\lambda}\,\partial^2\,\xi (x)\label{2.5}
\end{eqnarray}

In the spontaneously broken phase,  the theory is not any more  conformal but the % residual  trace 
memory of the  original symmetry is conserved in the infrared behaviour of the  S-matrix through the so called soft-theorems. These follow from the Ward-identities associated to the broken generators and assert that  amplitudes with one or more soft-dilatons are obtained acting  with suitable  operators related to the broken generators on amplitudes without the\mbox{ soft particles.   }

\subsection{Soft theorems from Ward-identities}
We consider a conformal invariant field theory whose conformal symmetry is spontaneously broken by the vev acquired by  some scalar field. In a such  theory we analyse the matrix element among the  divergences of set of broken currents and  a string of scalar fields
\begin{eqnarray}
\frac{\partial}{\partial y_1^{\mu_1} }\dots\frac{\partial}{\partial y_m^{\mu_m} } T^*\langle 0| j_1^{\mu_1} (y_1)\dots j_m^{\mu_m} (y_m) ~ \phi (x_1 )   \dots \phi (x_n) | 0\rangle \,,
\end{eqnarray}
$T^*$ denotes the usual Time-ordered product with all the derivative placed outside the time ordering \cite{coleman}. The derivatives of these matrix elements with respect the current coordinates determine the Ward-identities of the spontaneously broken conformal field theory.

In the case of a single  current, in momentum space,  one gets:
\begin{eqnarray}
&\int d^{d}x\,\text{e}^{-i q\cdot x}
\Bigl [
-\partial_\mu \,T^*\langle 0| j^{\mu} (x) \phi (x_1 ) \dots \phi (x_n) | 0\rangle
+
T^*\langle 0| \partial_\mu  j^{\mu} (x) \phi (x_1)   \dots \phi (x_n) | 0\rangle
\Bigr ]&\nonumber \\ 
&
=
-\sum_{i=1}^{n}\,\text{e}^{-i q\cdot x_{i}}
\,T^{*}\langle 0| \phi (x_1) \dots   \delta \phi (x_i)\dots \phi (x_n)| 0 \rangle\,.&\label{2.7}
\end{eqnarray}

The matrix element is evaluated up to the order ${\cal O}(q^0)$ in the momentum expansion  $q$ of the current. The first term on left side of this equation is neglected  after an integration  by part under the assumption of absence of poles at $q=0$. For the broken dilation and special conformal transformation,  Equation~\eqref{2.7} through  the divergences of the currents given in Equation~\eqref{2.5}     
relates the correlation functions with the insertion of a dilaton field with low momentum  to those without the dilaton but depending on the variations of the fields under the considered transformations. This identity contains soft-theorems at a very preliminary  stage and the LSZ-reduction translates them in a relation  between amplitudes. The LSZ-operator 
 \begin{align}
\Bigl[\text{LSZ}\Bigr] 
\equiv
i^{n}\left(\prod_{j=1}^{n}\lim_{k_{j}^{2}\rightarrow -m_{j}^{2}}\int d^{D}x_{j}\,\text{e}^{-i k_{j}\cdot x_{j}}(-\partial_{j}^{2}+m_{j}^{2})\right)\,,
%\label{LSZ}
\nonumber 
\end{align}
is applied on both sides of Equation~\eqref{2.7} with the on-shell limit  $k_j^2 \to - m_j^2$ performed only to the end of the calculation.  The current whose divergence gives the amplitude with   the  insertion of the dilaton,   is not truncated by the LZS-reduction formula.  The presence in the divergence of the current of the operator $-\partial^2$ amputates the dilaton propagator determining a relation between two truncated amplitudes.
This in the case of the dilatation current, up to the order ${\cal O}(q^0)$ in the soft-expansion,  gives:
\begin{eqnarray}
{v}\,
\mathcal{T}_{n+1}(q; k_{1},\ldots,k_{n})
=\Bigg \{
d - n\,\Delta - \sum_{i=1}^{n-1}k_{i}^{\mu}\frac{\partial}{\partial k_{i}^{\mu}}
- \sum_{i=1}^n \frac{m_i^2}{k_i \cdot q} 
\left ( 1 + q^\mu \frac{\partial}{\partial k_i^\mu} \right)
\Bigg\} 
\mathcal{T}_{n}(k_{1},\ldots, \bar{k}_n )\,.\nonumber\\
\label{2.8}
\end{eqnarray}
with $\bar{k}_n=-q+\sum_{i=1}^{n-1} k_i$.  Equation~\eqref{2.8} the amplitude with the insertion of a soft dilaton is obtained acting with the generator of the dilatation, written in the momentum space,  $\hat{\cal D} = - i k_{\mu} \frac{ \partial}{\partial k_\mu}$, on the amplitude without the dilaton. In the case of hard or finite energies  massless field,  the leading divergent contribution is vanishing.  For massive hard particles, instead, the presence of the pole in $q$ makes the leading soft-theorem of the dilaton very similar to that satisfied by the  gauge bosons of the fundamental interactions. More details on  such derivation are given in Refs. \cite{DMMN,1705.06175}, here it is worthwhile to comment that in Equation~\eqref{2.8} we have commuted the generator of the translations with the delta-function over the momenta and in the leading divergent term we have used the prescription to perform the on-shell limit before the soft one.

The same analysis is now performed for broken current generating the special conformal transformations.
The calculation is very similar to  the previous one with the only  difference that the operator of  the special conformal transformations commutes with the delta over the momenta and no extra factors, as the space-time dimension $d$ as in \mbox{Equation~\eqref{2.8},} is introduced.  The resulting Ward-identity is:
\begin{eqnarray}
&&{v}\,
q^{\lambda}
\frac{\partial}{\partial q^{\lambda}}
\mathcal{T}_{n+1}(q; k_{1},\ldots,-q-\sum_{j=1}^{n-1}k_{j})
\nonumber  \nonumber\\
=&&
\sum_{i=1}^{n}
\left\{
\frac{m_i^2}{k_{i}\cdot q} 
\left ( 1- \frac{1}{2} q^\mu q^\lambda \frac{\partial^2}{\partial k^\mu \partial k^\lambda} \right )
\right. \left. 
-q^{\lambda}\left[ k_i^{\mu} \left(\frac{\partial^2}{\partial k_{i}^{\mu} 
\partial k_{i}^{\lambda}} -
\frac{1}{2} \eta_{\mu\lambda}  \frac{\partial^2}{\partial k_{i\nu} \partial k_{i}^{\nu}}  
 \right)
+ d \frac{\partial}{\partial k_{i}^{\lambda} } \right]
\right\} 
\notag \\
&&\times 
\mathcal{T}_{n}(k_{1},\ldots,\bar{k}_n)+ \mathcal{O}(q^2) \,.
\label{KFinal}
\end{eqnarray}

The ${\cal O}(q^1)$ of the  amplitude with one dilaton and $n$-hard states $\mathcal{T}_{n+1}$ is obtained  by decomposition the amplitudes as the sum of the leading, subleading and subsubleading \mbox{soft terms:}
\begin{eqnarray}
\mathcal{T}_{n+1}
=&&
\left[
\sum_{i=1}^n\frac{\mathbf{S}_i^{(-1)}(q)}{k_{i}\cdot q}
+ \mathbf{S}^{(0)} + q^\mu \mathbf{S}^{(1)}_\mu\right]
\, \mathcal{T}_n(k_1, \ldots, k_n)
+ \mathcal{O}(q^2) \, ,
\end{eqnarray}
where the $\mathbf{S}^{(0)}$ and $\mathbf{S}_\mu^{(1)}$ are operators dependent only on the momenta $k_{j}$, while $\mathbf{S}^{(-1)}$ may depend on $q$ as well and replacing such expression in Equation~\eqref{KFinal}. The leading ${\cal O}(q^{-1})$ term cancels on both sides leading to an identity which determines:
\begin{eqnarray}
& v \mathbf{S}_i^{(-1)}\Big |_{\mathcal{O}(q^2)} =
- m_i^2 \left (
\frac{1}{2}q^\mu q^\lambda \frac{\partial^2}{\partial k^\mu \partial k^\lambda} \right)\,,
 \\
&v \mathbf{S}_\lambda^{(1)} 
 =
\sum_{i=1}^n \left[ \frac{k_{i\lambda}}{2} \frac{\partial^2}{\partial k_{i\nu} \partial k_{i}^{\nu}} 
-k_i^{\mu}\frac{\partial^2}{\partial k_{i}^{\mu} 
\partial k_{i}^{\lambda}}
- d \frac{\partial}{\partial k_{i}^{\lambda} } \right]
\end{eqnarray}

The Ward-identities of the scale and special conformal transformation determine through the subsubleading order the amplidude with one soft-dilaton in terms of the amplitude without the ditaton through the relation:

\begin{eqnarray}
\label{finalbehavior}
 &&{v}\,{\cal{T}}_{n+1} ( q; k_1, \ldots, k_n ) =
\Bigg\{ 
 - \sum_{i=1}^{n} \frac{m_i^2}{k_i \cdot q}
\left ( 1 + q^\mu \frac{\partial}{\partial k_i^{\mu}}
+ \frac{1}{2} q^\mu q^\nu \frac{\partial^2}{\partial k_i^\mu \partial k_i^{\nu} }\right)
+ d - n\Delta - \sum_{i=1}^{n}k_i^{\mu} \frac{\partial}{\partial k^{\mu}_i}\nonumber
 \\[.2cm] 
&&  \left.
-q^\lambda \sum_{i=1}^{n} \left[ 
\frac{1}{2}
\left(2\,k_i^{\mu} \frac{\partial^2}{\partial k_{i}^{\mu} 
\partial k_{i}^{\lambda}} -
k_{i\,\lambda}   \frac{\partial^2}{\partial k_{i\nu} \partial k_{i}^{\nu}}  
 \right)
+ \Delta\,\frac{\partial}{\partial k_{i}^{\lambda} } \right] 
 \right\}
\mathcal{T}_{n}(k_{1},\ldots,\bar{k}_n) + {\cal{O}} (q^2 )\,.
\end{eqnarray}

The terms proportional to the mass of the external states show a pole in the momentum of the soft particle, in perfect analogy with the the Low and Weinberg soft behaviours, and are determined by  the cubic couplings between a dilaton and two massive states. It may also be though as a $q$-expansion of the $n+1$-points  amplitude. The regular terms  are not vanishing  to the ${\cal O}(q^0)$ and are fixed by action of  broken generators, dilatation ${\cal D}=-i k_\mu\partial_{k_\mu}$ and special conformal transformations ${\cal K}_\mu =k_\mu \partial_{k}^2-2(k \cdot \partial_{k}) \partial_{k^\mu}$, on the amplitude without the emission at zero momentum of the dilaton.

The double soft behaviour of the amplitude with the emission of two dilatons at zero momenta is determined starting from the matrix elements with the insertion of  two broken currents and $n$-finite energy states. The two currents can be either of the same nature, dilatation or special conformal transformation, or  of different kind. Three sets of double-soft  Ward-identities may in principle be computed and they are related, through  LZS-reduction,   to the terms in the soft expansion of the amplitudes with the emission of two dilatons at zero momenta. This  calculation has only  been performed for external hard massless states and it is similar but more involved  
of the single soft identities. The leading ${\cal O}(q^0)$ term of the amplitude is obtained from the Ward-identities with two dilatation currents while the subleading behaviour is obtained from those with mixed broken generators. The subsubleading ${\cal O}(q^2)$ contribution of the amplitude would be determined by the Ward-identities with two special conformal transformation currents. These have not been possible to determine because as discussed  in Ref. \cite{1705.06175}, they would  require a formulation of the single soft-theorem to ${\cal O}(q^2)$ which is still missing.  
The universal behaviour of an amplitude with two dilatons at zero momenta is given in terms of the amplitude without the soft particles by the expression:

\begin{eqnarray}
&&v^2 T_{n+2}(q,k,k_1, \ldots, \bar{k}_n)
=
\Bigg [
\left ( d- \Delta_\xi - \sum_{i=1}^n \left(\Delta_i +k_i\cdot \partial_{k_i}\right)\right )\left ( d - \sum_{i=1}^n\left(\Delta_i+k_i\cdot \partial_{k_i}\right) \right )\nonumber
 \\
&&
+ (q_\mu + k_\mu) \sum_{i=1}^n\left(\frac{1}{2} k_i^\mu\partial_{k_i}^2 -\left(\Delta_i +k_i\cdot \partial_{k_i}\right) \partial_{k_i}^\mu\right) \left ( D - \Delta_\xi - \sum_{i=1}^n \left(\Delta_i+k_i\cdot \partial_{k_i}\right) \right )
\Bigg ]T_{n}( k_1, \ldots, \bar{k}_n) \nonumber\\
&&+ {\cal O}(q^2,k^2,qk)\nonumber
\label{fulldoublesoft}
\end{eqnarray}

It is straightforward to  verify that the double soft behaviour coincides with the one obtained by considering two consecutive single soft expansions. This observation has led to conjecture that the  behaviour of amplitudes with multiple  emission of  soft dilatons  is universal and it is  fixed  by the consecutive soft limits of amplitudes with  dilatons at zero momentum  emitted one after the other.

\section{String Amplitudes}
\label{String}
In string theories, amplitudes are very compact expressions given by complex integrals of correlation functions of vertex operators evaluated on Riemann surfaces of a given genus that represent
 the world-sheet of strings propagating in the space-time. 
At genus zero,  the sphere is the relevant tree-level closed string world-sheet and the  superstring  vertex operators  associated with the  massless states in the $(-1,\,-1)$ and $(0,0)$ super-ghost pictures take   the form \footnote{An introduction to superstring amplitudes can be found in the textbooks of   Refs. \cite{string1,string2}.}:
\begin{eqnarray}
V^{(-p,\,-p)}=\frac{\kappa_d}{2\pi}\int d\theta~\theta^p~ V^{(p)}(z,\,\theta;k) 
\int d\bar{\theta} ~\bar{\theta}^{p} ~  \bar{V}^{(p)}(\bar{z},\,\bar{\theta}; k) \, ,
\end{eqnarray} 
with $\kappa_d$ the gravitational coupling constant and  $z$ the complex coordinate parametrizing the insertion on the world-sheet of the vertex operators 
\begin{eqnarray}
V^{(p)}(z,\,\theta;k)=  e^{-p\phi(z)}~\epsilon_\mu~ DX^\mu\, 
e^{i\sqrt{\frac{\alpha'}{2}} k\cdot X(z,\,\theta)}
\, .
\end{eqnarray}

Here, $\theta$ and $\bar{\theta}$ are Grassmannian variables,
$\epsilon_\mu\bar{\epsilon}_\nu = \epsilon_{\mu\nu}$ is the polarization
 of the massless state, $k$ in the momentum of the external state,
and the superfield notation is given by
\begin{eqnarray}
 X^{\mu}  (z, \theta) &\equiv x^{\mu}(z) + \theta \psi^{\mu}(z) \, , \quad
D \equiv \frac{\partial}{\partial \theta} + \theta \frac{\partial}{\partial z} \, . 
%&& x^{\mu} (z) = {\hat{q}}^{\mu} - i {\hat{p}}^{\mu}\log z   +i  \sum_{n\in %\mathbb{Z}/\{0\}} \frac{\alpha_n^{\mu}}{n} z^{-n} ~~~;~~~
%\psi^{\mu} (z) = -i \sum_{r \in \mathbb{Z} + \frac{1}{2}} \psi_{r}^{\mu} z^{-r - %\frac{1}{2}}
\label{3}
\end{eqnarray}

This  vertex  describes collectively all $NS-NS$ massless states, these are the graviton $g_{\mu\nu}$, the antisymmetric two form $B_{\mu\nu}$ and the  scalar field (gravity dilaton) $\phi$ of the massless gravitational string multiplet. These fields are identified with  the  symmetric, anti-symmetric and trace of the external polarization, through the relation:
 \begin{eqnarray}
 \epsilon_\mu \bar{\epsilon}_\nu &&= 
\left [\frac{\epsilon_\mu \bar{\epsilon}_\nu + \epsilon_\nu \bar{\epsilon}_\mu}{2} 
- \frac{\epsilon_{\mu\nu}^\perp}{\sqrt{d-2}} \, \epsilon \cdot \bar{\epsilon}
\right ] 
%\nonumber \\
%&
+ \left[ \frac{\epsilon_{\mu\nu}^\perp}{\sqrt{d-2}} \,  
\epsilon \cdot \bar{\epsilon}
\right ] 
%\nonumber \\
%&
+\left [\frac{\epsilon_\mu \bar{\epsilon}_\nu - \epsilon_\nu \bar{\epsilon}_\mu}{2} 
\right ] 
\nonumber \\
&&=g_{\mu \nu}(k) + \frac{\epsilon_{\mu\nu}^\perp}{\sqrt{d-2}}  \phi(k) + B_{\mu \nu}(k)\label{3.19}
\end{eqnarray}
where
\begin{eqnarray}
\epsilon_{\mu\nu}^\perp = \frac{\eta_{\mu \nu} - k_\mu \bar{k}_\nu-k_\nu \bar{k}_\mu}{\sqrt{d-2}}
\, , \quad
k^2 = \bar{k}^2=0 \, , \quad k\cdot \bar{k} = 1
\end{eqnarray}
such that $\eta^{\mu \nu} \epsilon_{\mu \nu}^\perp = \sqrt{d-2}$, and
$\epsilon_{\mu \nu}^\perp {\epsilon^\perp}^{\mu \nu} = 1$, while
$\bar{k}$ is an unphysical reference momentum.
The relevant expectation values for the world-sheet 
fields are:
\begin{eqnarray}
\langle X^{\mu} (z_1, \theta_1) X^{\nu} (z_2, \theta_2) \rangle &&= - 
\eta^{\mu \nu} \log(z_1 -z_2 - \theta_1 \theta_2) \, , \\
\langle e^{-\phi(z_1)}\,e^{-\phi(z_2)}\rangle &&=\frac{1}{z_1-z_2} \, .
\label{5}
\end{eqnarray}

The superstring amplitude with $n+1$ massless external states is:
\begin{eqnarray}
M_{n+1} =&&
\frac{8\pi}{\alpha'}\left (\frac{\kappa_d}{2\pi}\right )^{n-1}
 \int \frac{d^2 z \prod_{i=1}^{n} d^2 z_i d\theta d\bar{\theta} }{dV_{abc} 
|z_1  - z_2|^2}\bigg[\prod_{i=1}^2d\theta_i\theta_i \prod_{i=3}^{n} 
d \theta_i\bigg]\bigg[\prod_{i=1}^2d\bar{\theta}_i\bar{\theta}_i 
\prod_{i=3}^{n} d {\bar{\theta}}_i\bigg]  \nonumber \\
&& \times  \langle 0 | 
\int d \varphi \,\,e^{ i \left( \varphi \epsilon  D X (z, \theta) + 
\sqrt{\frac{\alpha'}{2}}q  X (z, \theta)  \right)}
\prod_{i=1}^{n} \left( \int d \varphi_i  \,\,e^{ i \left( \varphi_i 
\epsilon_i  D_i X (z_i, \theta_i) + K_i  X (z_i, \theta_i)  \right)} \right) 
| 0\rangle \nonumber \\
&&  \times  \langle 0 | 
\int d {\bar{\varphi}} \,\,e^{ i \left( {\bar{\varphi}} {\bar{\epsilon}} 
 {\bar{D}} X ({\bar{z}}, {\bar{\theta}}) + \sqrt{\frac{\alpha'}{2}}q  
X ({\bar{z}}, {\bar{\theta}})  \right)}
\prod_{i=1}^{n} \left( \int d {\bar{\varphi}}_i  \,\,e^{ i 
\left( {\bar{\varphi}}_i {\bar{\epsilon}}_i  {\bar{D}}_i X ({\bar{z}}_i, {\bar{\theta}}_i) + K_i  X ({\bar{z}}_i, {\bar{\theta}}_i)  \right)} \right) | 0\rangle \,\nonumber\\ .
\label{super1}
\end{eqnarray}

Here, $dV_{abc}$ is the volume of the M\"obius group,  the states with the indices $1$ and $2$ are in the $(-1, -1)$ picture, while the others are in the $(0,0)$ picture and the factor $\prod_{i=1}^2 
\theta_i \bar{\theta}_i/|z_1-z_2|^2$ comes from the correlator of 
the superghosts.

Equation~\eqref{super1} is the starting point for computing  amplitudes with $n+1$ external massless states. In such an expression, one vertex operator depending on the complex super-coordinate $Z=(z,\,\theta,\,\bar{\theta})$ and  momentum $q$ is kept  separated from all the other external legs. It will be considered the vertex of the soft particle whose momentum will be expanded up to order  ${\cal O}(q)$.  
Soft theorems will be obtained by integrating over the super-coordinate $Z$  and expressing the result of the integration by an operator acting on the amplitude without the soft-leg. This operator, in general, contains $\alpha'$  string corrections, being $\alpha'$ the Regge slope. The leading order in $\alpha'$ gives the correct field theory result.  Furthermore, 
%as seen from 
Equation~\eqref{super1} 
provides in a unique expression  the amplitudes for the 
graviton, the Kalb-Ramond and the
gravity dilaton. 
The analyses of the IR properties of these amplitudes give, in a single expression, the soft theorems of these three different particles. In the next section,  we will review those of the gravitational dilaton highlighting the similarities and differences with those of the conformal dilaton.

\subsection{Soft-dilaton behaviour}
The analyses of the infrared properties of amplitudes with Gravitational dilatons have a long history. They were already studied in the old dual models in connection with the renormalization of the Regge slope \cite{Ademollo:1975pf} and in the framework of closed string field \mbox{theories \cite{Hata:1992}.}
In this section we review the recent extension of these old results by studying the soft properties of amplitudes with scalar particles. The starting point of this analysis is Equation~\eqref{super1} supplemented by the observation that the amplitude after having integrated over the Grassmann variables associated to the soft vertex reduces to an expression that is  factorized at integrand level:
\begin{eqnarray}
M_{N+1} = M_n\star S
\end{eqnarray}
where the $\star$ denotes a convolution integral between the $n$-point amplitude $M_n$ and the integral $S$  which collects all the dependence %of the amplitude 
from the soft-leg.  This quantity
 is the sum of three terms $S=S_b+S_s+\bar{S}_s$ where $S_b$  receives 
contributions form the purely bosonic degrees of freedom and it is equal to the similar expression computed in bosonic string theory while $S_s$ and its complex conjugate $\bar{S}_s$ receive contribution from the supersymmetric partners of the bosonic degrees of freedom. This decomposition turns out to be very useful because from $S_b$ one can deduce the soft function in bosonic string. These %expressions 
have been computed in Ref. \cite{DiVecchia:2016szw}, here  we only cite their expressions. The bosonic contribution turns out to be:
 \begin{eqnarray}
 S_b= &\,\frac{\kappa_d}{2\pi} \int d^2 z \prod_{l=1}^{n}
 | z - z_l |^{ \alpha' q k_l} 
\,\prod_{l=1}^{n} {\rm exp}\left [ -\sqrt{\frac{\alpha'}{2}}\frac{q \cdot C_l}{z-z_l}-  
\sqrt{\frac{\alpha'}{2}}\frac{q \cdot \bar{C}_l}{{\bar{z}}-{\bar{z}}_l} \right] 
 \\
& \times 
\left( \sum_{i=1}^{n} \frac{\epsilon \cdot C_i}{(z-z_i)^2} +  
\sum_{i=1}^{n}\sqrt{\frac{\alpha'}{2}} \frac{\epsilon \cdot k_i}{z-z_i} \right)
%\nonumber \\
%& \times 
\left( \sum_{j=1}^{n} \frac{\bar{\epsilon} \cdot 
{\bar{C}}_j}{({\bar{z}}-{\bar{z}}_j)^2} + \sum_{j=1}^{n}\sqrt{\frac{\alpha'}{2}} 
\frac{{\bar{\epsilon}} \cdot k_j}{{\bar{z}}-{\bar{z}}_j} \right)
\, ,
\end{eqnarray}

The contribution dues to the superpartner degrees of freedom is:
\begin{eqnarray}
&&\bar{S}_s=\frac{\kappa_d}{2\pi} \int d^2 z\prod_{l=1}^{n}
 | z - z_l |^{ \alpha' q k_l} 
\,\prod_{l=1}^{n}  {\rm exp}\left [ -\sqrt{\frac{\alpha'}{2}}\frac{q \cdot C_l}{z-z_l}-  
\sqrt{\frac{\alpha'}{2}}\frac{q \cdot \bar{C}_l}{{\bar{z}}-{\bar{z}}_l} \right]  \\
 &&\times\left[ \frac{1}{2}\sum_{i=1}^{n} \sqrt{\frac{\alpha'}{2}}
\frac{q \cdot A_i}{z-z_i} \sum_{j=1}^{n} \frac{\epsilon \cdot A_j}{z-z_j}
 \sum_{l=1}^{n}\sqrt{\frac{\alpha'}{2}} \frac{q \cdot
 {\bar{A}}_l}{{\bar{z}}-{\bar{z}}_l} \sum_{m=1}^{n} \frac{\bar{\epsilon} 
\cdot{\bar{A}}_m}{{\bar{z}}-{\bar{z}}_m} \right.
 \\
 &&+\left. \left (\sum_{i=1}^{n} \frac{\epsilon \cdot C_i}{(z-z_i)^2} +  
\sum_{i=1}^{n}\sqrt{\frac{\alpha'}{2}} \frac{\epsilon \cdot k_i}{z-z_i} 
\right ) \sum_{j=1}^{n}\sqrt{\frac{\alpha'}{2}} \frac{q 
\cdot {\bar{A}}_j}{{\bar{z}}-{\bar{z}}_j} \sum_{l=1}^{n} \frac{\bar{\epsilon} 
\cdot{\bar{A}}_l}{{\bar{z}}-{\bar{z}}_l} 
  \right] \, ,
\end{eqnarray}

The~superkinetic %MDPI: is the italic necessary?.
 quantities introduced in these expressions are given by:
\begin{eqnarray}
 A_i^\mu = \varphi_i \epsilon_i^\mu  +\sqrt{\frac{\alpha'}{2}} \theta_i k_i^\mu
~~;
 ~~ C_i^\mu = \varphi_i \theta_i  \epsilon_i^\mu \, ,\label{1.4}
\end{eqnarray}
with $\varphi$  Grassmannian variables introduced to exponentiate   the vertex operators. All the integrals in $S_b$ and $S_S$ have been computed in Refs.~\cite{DMM1,DMM2,DiVecchia:2016szw} up the ${\cal O}(q)$ in the expansion of the momentum of the graviton chosen to be the soft external leg. The result of the integrations, when the soft leg  is projected on the external symmetric and traceless polarization,    reproduces, according to Equation~\eqref{3.19}, the known soft theorems for gravitons. The projection on the dilaton state, instead,  is written  as the sum of  two operators acting on the amplitude depending only on hard momenta: 
\begin{eqnarray}
M_{n+1}^{\rm dilaton} = &&\,
\frac{\kappa_d}{\sqrt{d-2}} \left[ 2 - \sum_{i=1}^n k_{i\mu} 
\frac{\partial}{\partial k_{i\mu}} 
\right .  \\
&& \left. + \frac{1}{2}  \sum_{i=1}^n
\left( q^\rho {\hat{K}}_{i \rho} 
 + \frac{q^\rho q^\sigma}{k_i q}
\left(
\mathcal{S}_{i, \rho \mu}\eta^{\mu \nu} \mathcal{S}_{i \nu \sigma} + D
\Pi_{i,\{\rho, \sigma\}} \right) 
\right)  \right] M_n \, , 
\label{dilasoftfin}
\end{eqnarray}
where 
\begin{eqnarray*}
&&  {\hat{K}}_{i\mu} = 2 \left[ \frac{1}{2} k_{i \mu} \frac{\partial^2}{\partial
k_{i\nu} \partial k_i^\nu} 
-k_{i}^{\rho} \frac{\partial^2}{\partial
k_i^\mu \partial k_{i}^{\rho}} 
+ i \mathcal{S}_{i,\rho \mu} \frac{\partial}{\partial k_i^\rho}  \right]~~;~~\Pi_i^{\rho \sigma} = 
\epsilon_i^\rho\frac{\partial }{\partial \epsilon_{i\sigma}} +
{\bar{\epsilon}}_i^\rho\frac{\partial }{\partial {\bar{\epsilon}}_{i\sigma}} .
\label{hatDhatKmu1}
\end{eqnarray*}
with $\mathcal{S}_{\rho \mu} $ the spin-operator. Remarkably these operators are the generators of the dilatations and special conformal transformations acting in the momentum space. This result has also been confirmed by decomposing the $n+1$-amplitude in a double copy three-point vertex, with a soft leg  describing collectively a graviton,  a dilaton and a Kalb-Ramond,  factorized  from an $n$-point amplitude \cite{DiVecchia:2017gfi}. The expansion of this amplitude in the soft momentum  and the request of the gauge invariance imposed order by order in soft expansion, reproduces the dilaton infrared behaviour obtained by the explicit calculation of superstring amplitudes.  
The same calculation has also been performed in heterotic and bosonic string theories for massless and massive external state. Differently from the soft graviton behaviour which is universal up to subleading order, the tree-level soft behaviour of the dilaton is universal through the  ${\cal O}(q)$ in all string theories. It doesn't contain any string correction and therefore is a result that can also be obtained from amplitudes computed in the framework of  the  low energy effective string theories.
This result has been extended at $h$-order in the loop expansion by computing in bosonic string theory amplitudes with one massless state in interaction with $n$-tachyon states treated as massive particles with mass $m^2=-\frac{1}{\alpha'}$.  The calculation has been performed in space-time dimensions bigger than four to avoid infrared logarithmic divergences, the resulting soft-expansion reads:

\begin{eqnarray}
M_{n;\phi}^{(1)} (k_i; q) = 
\frac{\kappa_d}{ \sqrt{d-2} }  \left[ - \sum_{i=1}^n \frac{m^2}{k_i q} {\rm e}^{q \partial_{k_i}}  +2+h(d-2)- \sum_{i=1}^n k_i\cdot \frac{\partial}{\partial k_i}
+  q_\mu \sum_{i=1}^n {\hat{K}}_{i}^{ \mu}
   \right]  M_n^{(1)}    + {\cal O}(q^2) \,.\nonumber\\
\end{eqnarray}

Remarkably,   the multiloop soft expansion is again free by string corrections to any order in the loop expansion. Instead, since the multiloop infrared behaviour has been obtained only in the bosonic string, it is not clear if universality holds to any loop.

In analogy with the dilaton of the broken conformal field  theories, also the gravitational  dilaton obeys a double soft factorization property. This has been determined by computing in bosonic string theory tree level amplidudes with $n$-tachyons and two massless states carrying low momenta. The simultaneos  zero momentum limit is performed by rescaling, by a small parameter $\tau$, 
the momenta of two massless states and expanding the amplitude in $\tau$. The projection on the dilatons, up to the subleading order,  gives:
\begin{eqnarray}
M_{n+d+d} =
%\left (
\frac{\kappa_d^2}{d-2 } 
%\right ) 
\Bigg [&&
\frac{1}{\tau^2} \sum_{i,j=1}^n\frac{m_i^2\,m_j^2}{k_iq\, k_jl} 
\left( 1+ \tau q\frac{\partial}{\partial k_i}+\tau l\frac{\partial}{\partial k_j}\right)
-\frac{1}{\tau} 
\sum_{i=1}^n \frac{m_i^4}{(k_il)(k_iq) } \frac{(ql)}{k_i(q+l)}
\nonumber\\
&&
- \frac{d-2}{\tau} \sum_{i=1}^n \frac{(qk_i)(lk_i)}{ k_i(q+l)\,(ql)} 
-\frac{1}{\tau} 
\sum_{i=1}^n \frac{2 m_i^2}{k_i(q+l)} 
\nonumber\\
&&
-\frac{1}{\tau}\sum_{i=1}^n\left(\frac{m_i^2}{ k_iq}+\frac{m_i^2}{ k_il}\right)
\left ( 2-\sum_{j=1}^n k_j\frac{\partial }{\partial k_j} \right )
\Bigg]M_n  + {\cal O}(\tau^0)
 \label{3.30}
\end{eqnarray}

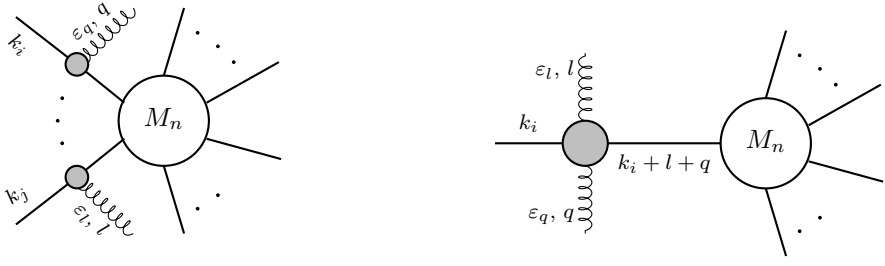
\begin{figure}[t]
%\begin{center}
\begin{tikzpicture}[scale=.30]
		\draw [thick]  (8.5,2) -- (9.4,5); 
		\draw [thick]  (10.4,0.8) -- (13.6,2.6); 
		\draw [thick]  (6.75, 0.8) -- (4.65,2.5);% (intermediate leg)
		\draw [thick]  (4.65, 2.5) -- (1.95,4.6) node[near end, sloped,below] { \scriptsize $k_{i}$};
		\draw [thick]  (4.65, -2.5) -- (1.95,-4.6) node[near end, sloped, above] { \scriptsize $k_{j}$};
		\draw [thick]  (6.75,-0.8) -- (4.65,-2.5); % (intermediate leg)
		\draw [thick]  (10.4,-0.8) -- (13.7,-1.7); 
		\draw [thick]  (8.5,-2) -- (9.4,-5);
		\draw[snake=coil,segment length=4pt] (7.15,5) -- (4.65,2.5) node[midway,sloped,above] { \scriptsize $\varepsilon_q,\, q$};
		\filldraw [thick, fill=lightgray] (4.65,2.5) circle (0.5cm);
		\draw [snake=coil,segment length=4pt] (7.15, -5) -- (4.65,-2.5) node[midway,sloped,below] { \scriptsize $\varepsilon_l,\, l$};
		\filldraw [thick, fill=lightgray] (4.65,-2.5) circle (0.5cm);
		\draw (8.5,.1) node { \footnotesize $M_n$};
		\draw [thick] (8.5,0) circle (2cm);
		\begin{scope}[shift={(-3,0)}] 
		\filldraw [ thick] (13.0,3.9) circle (1pt);
		\filldraw [ thick] (13.9,3.3) circle (1pt);
		\filldraw [ thick] (14.6,2.6) circle (1pt);
		\filldraw [ thick] (13.0,-3.9) circle (1pt);
		\filldraw [ thick] (13.9,-3.3) circle (1pt);
		\filldraw [thick] (7,1) circle(1pt);
		\filldraw [thick] (6.8,0) circle(1pt); 
		\filldraw [thick] (7,-1) circle(1pt); 
		\draw (10,-7.5) node { \footnotesize};
		\end{scope}
		\end{tikzpicture}
		\hspace{1.0in}	
		\begin{tikzpicture}[scale=.30]
		\draw [thick]  (8.5,2) -- (9.4,5); 
		\draw [thick]  (10.4,0.8) -- (13.6,2.6); 
		\draw [thick]  (10.4,-0.8) -- (13.7,-1.7); 
		\draw [thick]  (8.5,-2) -- (9.4,-5); 
		\draw [thick] (-3.5,0) -- (-0.5,0) node[midway, above]{ \scriptsize $k_i$};
		\draw [thick](1.5,0) -- (6.5,0)node[midway, below] {\scriptsize $k_i+l+q$}; % (intermediate leg)
		\draw [snake=coil,segment length=4pt](0.5,1) --(0.5,4) node[near end, left]{\scriptsize $\varepsilon_l,\,l$};
		\draw[snake=coil,segment length=4pt] (0.5,-1) -- (0.5,-4) node[near end, left] { \scriptsize $\varepsilon_q,\, q$};
		\filldraw [thick, fill=lightgray] (0.5,0) circle (1cm);
		\draw (8.5,.1) node { \footnotesize $M_n$};
		\draw [thick] (8.5,0) circle (2cm);
		\begin{scope}[shift={(-3,0)}] 
		\filldraw [ thick] (13.0,3.9) circle (1pt);
		\filldraw [ thick] (13.9,3.3) circle (1pt);
		\filldraw [ thick] (14.6,2.6) circle (1pt);
		\filldraw [ thick] (13.0,-3.9) circle (1pt);
		\filldraw [ thick] (13.9,-3.3) circle (1pt);
		\draw (7.5,-6.5) node { \footnotesize };
		\end{scope}
		\end{tikzpicture}
%\end{center}
\caption{Diagrams of two massless closed  string states with polarization and momenta $\varepsilon_q, q$ respectively and  $\varepsilon_l, l$. External tachyon lines have momenta $k_i$, with $i=1, \ldots, n$.
}
\label{Fig1}
\end{figure}

This result is independent of the string slope $\alpha'$ and it has  also been obtained by starting from Feynman diagrams computed in the low energy effective string theory. Two classes  of exchange Feynman diagrams contribute to
  the infrared behaviour of the  amplitudes 
(see Figure \ref{Fig1}). The first class of diagrams involve only three-point vertices with two tachyons and one dilaton emitted from two different external lines. In the second class of diagrams, three and four-point vertices are emitted from a single external line. These are the  vertices with two tachyons and one massless state, three massless states and a four point-vertex with two tachyons and two massless states.  
The contribution due to the four-point interaction is collected by the single pole term in second line in Equation~\eqref{3.30}.  Gauge invariance is  obtained by requiring the emission of the massless states from internal lines. The same expression is also obtained by joining a four-point amplitude, with two tachyons and two dilatons, to a $n$-point  tachyon amplitude. It is straightforward to see that Equation~\eqref{3.30} differs from the expression obtained by considering two consecutive single soft expansions in the term generated by the four-point interaction. 

\section{Discussions}
In this paper, we have summarized the state of the art of the soft-theorems of two different particles named dilaton, the Nambu-Goldstone boson of the spontaneously broken conformal theories and the scalar of the massless gravitational multiplet of the string theories.  Symmetries of the theory determine the single and double soft behaviour for both particles.  Ward-identities associated with the scale and special conformal transformation control the soft-behaviour of amplitudes with conformal dilatons.  Gauge-invariance of string (or string effective theory) amplitudes fixes the infrared behaviour of all the massless closed string multiplet.
The single soft theorems for both particles are almost similar.  They contain at subleading order the generators of the scale transformations and kinematically invariant factors related to the mass and scale dimensions of the amplitudes. The subsubleading order in both cases is fixed by the generators of the special conformal transformations. 
These analogies  are less evident in the double soft behaviour of the two dilatons. Both contain the generators of the dilations,  but for external massless finite energy particles, the amplitudes with conformal soft dilatons are finite at zero momentum while diverging for the gravity dilaton. Furthermore,  the simultaneously low momentum limit for the conformal dilatons coincides with the consecutive one. That is not true in the case of the gravity dilaton for the presence of quartic interaction among two massive and two dilaton particles.
Differences appear also at higher order in the perturbative expansion.  
In string theory, the dilaton remains massless to all loops and satisfies factorization properties similar to the tree-level one. In conformal theories, the dilaton, in general,  become massive at one-loop  because of the breaking of conformal invariance in quantum theories.  
We emphasize   that  it is still missing a  symmetry principle that explains the infrared properties of the gravitational dilaton. 
These are inferred  by direct amplitude calculations or by the request of gauge invariance of string amplitudes that describe in a unified expression scattering of gravitons, Kald-Ramonds and dilatons. 
Similarities between the soft theorems of the two dilatons suggest that also the IR properties of the gravity-dilaton might be determined by the breaking of some ``hidden'' symmetry of the theory.

 Universality of the loop soft theorem for the gravity-dilaton is still an open issue.  The loop factorization properties for this kind of particle have been derived only in bosonic string theory and for external tachyons treated as massive scalar fields \cite{Scherk:1971}. They hold also in field theory because the soft operators are independent on the string slope  but a their extension  to other string models and for other kinds of hard states is still lacking. 
 
 We conclude these notes by observing that the gravity dilaton is a massless scalar that appears in all superstring theories and can play an important role in the so-called string cosmology. They have been proposed, for example,  cosmological models where the primordial evolution of the background is driven by this massless scalar.
On the other side, there are extensions of the Standard Model and gravity based on scale invariance which could solve the hierarchy problem. In all these cosmological contests it would be interesting to explore the role played by soft theorems in establishing relations among \mbox{cosmological correlators.}

%%%%%%%%%%%%%%%%%%%%%%%%%%%%%%%%%%%%%%%%%%

%%%%%%%%%%%%%%%%%%%%%%%%%%%%%%%%%%%%%%%%%%

%%%%%%%%%%%%%%%%%%%%%%%%%%%%%%%%%%%%%%%%%%

{\bf Acknowledgement:} 
 I would like to  thank   Paolo Di Vecchia, Matin Mojaza,  Josh Nole and Mritunjay Verma for having accompanied me on this journey into the formulation of soft theorems for dilatons.
I thank P. Di Vecchia for his helpful comments on the manuscript.

%%%%%%%%%%%%%%%%%%%%%%%%%%%%%%%%%%%%%%%%%%

%%%%%%%%%%%%%%%%%%%%%%%%%%%%%%%%%%%%%%%%%%


\begin{thebibliography}{999}
\bibitem{Low1}
Low, F.E. %MDPI: there are not only one ref information in this ref, please separted into 5 refs and change the ref citation, please confirm and revise.
{ Scattering of light of very low frequency by systems of spin 1/2}. 
\emph{Phys.\ Rev}. \textbf{1954}, { \emph{96}}, 1428.
%%CITATION = PHRVA,96,1428;%%

\bibitem{Low2}
Low, F.E.
{ Bremsstrahlung of very low-energy quanta in elementary particle collisions}.
\emph{Phys.\ Rev.\ } \textbf{1958}, {\emph{110}}, 974.
%%CITATION = PHRVA,110,974;%%
%

\bibitem{Low3}
Gell-Mann, M.; Goldberger, M.L.
{ Scattering of low-energy photons by particles of spin 1/2}.
\emph{Phys.\ Rev.} \textbf{1954}, {\emph{96}}, 1433.
%%CITATION = PHRVA,96,1433;%%
%

\bibitem{Low4}
Saito, S.
{ Low-energy theorem for Compton scattering}. 
\emph{Phys.\ Rev. } \textbf{1969},   {\emph{184}}, 1894.
%%CITATION = PHRVA,184,1894;%%
%+% 5 refs

\bibitem{Low5}
 Low, F.E.
{ Bremsstrahlung of very low-energy quanta in elementary particle collisions}. 
\emph{Phys. Rev.}  \textbf{1958}, { \emph{110}}, 974.
%%CITATION = PHRVA,110,974;%%

\bibitem{Weinberg1}
Weinberg, S. %MDPI:  there are not only one ref information in this ref, please separted into 3 refs and change the ref citation, please confirm and revise.
{ Photons and Gravitons in s Matrix Theory: Derivation of Charge Conservation and Equality of Gravitational and Inertial Mass}. 
\emph{Phys. Rev.}  \textbf{1964}, { \emph{135}}, B1049.
%%CITATION = PHRVA,135,B1049;%%
%
%S.~Weinberg,
%Phys.\ Rev.\  {\bf 140}, B516 (1965).
%%CITATION = PHRVA,140,B516;%%

\bibitem{Weinberg2}
Gross, D.J.; Jackiw, R.
{ Low-Energy Theorem for Graviton Scattering}. 
\emph{Phys. Rev.}  \textbf{1968}, {\emph{166}}, 1287.
%%CITATION = PHRVA,166,1287;%%

\bibitem{Weinberg3}
Jackiw, R.
{ Low-Energy Theorems for Massless Bosons: Photons and Gravitons}. 
\emph{ Phys. Rev.} \textbf{1968},  {\emph{168}}, 1623.
%+% 1 ref
%+% 4 refs 

\bibitem{gauge1}
Broedel, J.; de Leeuw, M.; Plefka, J.; Rosso, M. %MDPI: please separate the ref into 2 refs, please confirm and revise.
 {Constraining subleading soft gluon and graviton theorems}.  \emph{Phys. Rev.} \textbf{2014}, { \emph{D90}},  065024. %MDPI: delete arxiv number, please confirm
%%CITATION = ARXIV:1406.6574;%%

\bibitem{gauge2}
Bern, Z.; Davies, S.; Vecchia, P.D.; Nohle, J, 
{Low-Energy Behavior of Gluons and Gravitons from Gauge Invariance}. 
\emph{Phys. Rev.} \textbf{2014}, \emph{D90}, 084035.
%P. Di Vecchia, R. Marotta, M. Mojaza, {\em 
%The B-field soft theorem and its unification with the graviton and dilaton}, JHEP 10 (2017) 017, 
%arXiv: 1706.02961 [hep-th].
\bibitem{DiVecchia:2016szw} Vecchia, P.D.; Marotta, R.; Mojaza, M. { Soft behavior of a closed massless state in superstring and universality in the soft behavior of the dilaton}. \emph{J. High Energy Phys.} \textbf{2016}, \emph{12}, 020.

\bibitem{1611.07534}
Elvang, H.; Jones, C.R.T.; Naculich, S.G. { Soft Photon and Graviton Theorems in Effective Field Theory}. \emph{Phys. Rev. Lett.}  \textbf{2017},  \emph{118}, 231601. 



\bibitem{uni1}
%\bibitem{1404.4091} 
 Cachazo, F.; Strominger, A. %MDPI:there are not only one ref information in this ref, please separted into 3 refs and change the ref citation, please confirm and revise.
  { Evidence for a New Soft Graviton Theorem}. \emph{arXiv} \textbf{2014},
  arXiv:1404.4091.
  %%CITATION = ARXIV:1404.4091;%%
%\bibitem{1404.7749} 

\bibitem{uni2}
 Schwab, B.U.W.; Volovich, A.
  { Subleading Soft Theorem in Arbitrary Dimensions from Scattering Equations}. 
\emph{  Phys. Rev. Lett. } \textbf{2014}, \emph{113}, 101601. 
 % doi:10.1103/PhysRevLett.113.101601
  %%CITATION = doi:10.1103/PhysRevLett.113.101601;%%
%\bibitem{1405.1410} 

\bibitem{uni3}
  He, S.; Huang, Y.t.; Wen, C.
  {Loop Corrections to Soft Theorems in Gauge Theories and Gravity}. 
  \emph{J. High Energy Phys.} \textbf{2014},  \emph{1412}, 115.
  %%CITATION = doi:10.1007/JHEP12(2014)115;%%
%\bibitem{1405.2346} 

\bibitem{uni4}
  Larkoski, A.J.
  {Conformal Invariance of the Subleading Soft Theorem in Gauge Theory}.
\emph{  Phys. Rev. D } \textbf{2014}, \emph{90},  087701.
  %%CITATION = doi:10.1103/PhysRevD.90.087701;%%  
%\bibitem{Bianchi:2014gla}

\bibitem{uni5}
Bianchi, M.; He, S.; Huang, Y.T.; Wen, C. {More on Soft Theorems: Trees, Loops and Strings}.
\emph{ Phys. Rev. D } \textbf{2015}, \emph{92}, 065022.
%\bibitem{1406.7184} 

\bibitem{uni6}
 White, C.D.
  { Diagrammatic insights into next-to-soft corrections}, 
  \emph{Phys. Lett. B} \textbf{2014},  \emph{737}, 216.
  %%CITATION = doi:10.1016/j.physletb.2014.08.041;%%  
%\bibitem{1407.5936} 

\bibitem{uni7}
  Zlotnikov, M.
  { Sub-sub-leading soft-graviton theorem in arbitrary dimension}. 
  \emph{J. High Energy Phys.} \textbf{2014}, {\emph{1410}}, 148.
  %%CITATION = doi:10.1007/JHEP10(2014)148;%%    
  %\bibitem{1407.5982}
  
  \bibitem{uni8}
  Kalousios, C.; Rojas, F. 
  { Next to subleading soft-graviton theorem in arbitrary dimensions}, 
  \emph{J. High Energy Phys.} \textbf{2015}, {\emph{1501}}, 107.
  
  \bibitem{uni8b}
 % \bibitem{1408.4179} 
  Du, Y.J.; Feng, B.; Fu, C.H.; Wang, Y.
  { Note on Soft Graviton theorem by KLT Relation}.
  \emph{J. High Energy Phys.} \textbf{2014}, {\emph{1411}}, 090.
  %%CITATION = doi:10.1007/JHEP11(2014)090;%%
%\bibitem{Schwab:2014sla}

\bibitem{uni9}
Schwab, B.U.W. {A Note on Soft Factors for Closed String Scattering}.  \emph{J. High Energy Phys.} \textbf{2015},  {\emph{1503}}, 140.
%\bibitem{1412.3699}

\bibitem{uni10}
 Vera, A.S.; Vazquez-Mozo, M.A.
  {The Double Copy Structure of Soft Gravitons}. 
  \emph{J. High Energy Phys.} \textbf{2015}, \emph{1503}, 070.
  %%CITATION = doi:10.1007/JHEP03(2015)070;%%
%\cite{Guerrieri:2015eea}
%\bibitem{Guerrieri:201eea}

\bibitem{uni11}
Guerrieri, A.L. { Soft behavior of string amplitudes with external massive states}. 
\emph{Nuovo Cim. C} \textbf{2016}, {\emph{39}},  221. 
%\bibitem{DiVecchia:2015srk}

\bibitem{uni12}
Vecchia, P.D.; Marotta, R.; Mojaza, M. {Soft Theorems from String  Theory}. \emph{Fortschr. Phys.} \textbf{2016}, { \emph{64}}, 389.

\bibitem{uni13}
%\bibitem{Bianchi:1512}
Bianchi, M.; Guerrieri, A.L.  { On the soft limit of closed string amplitudes with massive states}. 
\emph{ Nucl. Phys. B} \textbf{2016}, {\emph{905}},  188.
%  %%CITATION = doi:10.1016/j.nuclphysb.2016.02.005;%%

\bibitem{uni14}
%\cite{Sen:2017xjn}
%\bibitem{Sen:2017xjn}
Sen, A. { Soft Theorems in Superstring Theory}. 
 \emph{J. High Energy Phys.} \textbf{2015},  {\emph{1706}}, 113.
 
\bibitem{uni15}
%%
Sen, A. { Subleading Soft Graviton Theorem for Loop Amplitudes}.  \emph{J. High Energy Phys.} \textbf{2017}, \emph{11}, 123.

\bibitem{uni16}
%https://doi.org/10.1007/JHEP11(2017)123
Laddha, A.; Sen, A. {Sub-subleading soft graviton theorem in generic theories of quantum gravity}. 
\emph{J. High Energy Phys.} \textbf{2017}, \emph{11}, 65.
%https://doi.org/10.1007/JHEP10(2017)065

\bibitem{uni17}
%\bibitem{Higuchi:2018vyu}
Higuchi, S.; Kawai, H. { Universality of soft theorem from locality of soft vertex operators}. 
\emph{Nucl. Phys. B} \textbf{2018}, {\emph{936}}, 400--447. 

\bibitem{uni18}
Bhatkar, S.A.; Sahoo, B. { Subleading Soft Theorem for arbitrary number of external soft photons and gravitons}. \emph{J. High Energy Phys.} \textbf{2019}, \emph{01}, 153.
%https://doi.org/10.1007/JHEP01(2019)153
%%%%%%%%%%%%%%%%%%%%%%%%

%%%%%%%%%%%%%%%%%%%%%%%%%%%
\bibitem{1405.1015}
Bern, Z.; Davies, S.; Nohle, J. {One Loop Corrections to Subleading Soft Behaviour of Gluons nd Gravitons}. \emph{Phys. Rev. D} \textbf{2014}, \emph{90}, 085015.
%DOI:https://doi.org/10.1103/PhysRevD.90.085015
 
%%%%%%%%%%%%%%%%%%%%%%%%%%%%%%%%%%%%%%%%%%%%%%%%%%%%%%%%%%%%%
\bibitem{Weinberg:Photon}
Weinberg, S. %MDPI: there are not only one ref information in this ref, please separted into 3 refs and change the ref citation, please confirm and revise..
 { Infrared Photons and Gravitons}. \emph{Phys. Rew.} \textbf{1965}, { \emph{140}}, B516.
%%DOI:https://doi.org/10.1103/PhysRev.140.B516
 
 \bibitem{log1}

%\bibitem{Laddha:2018myi} 
  Laddha, A.; Sen, A. {Logarithmic Terms in the Soft Expansion in Four Dimensions}. 
  \emph{J. High Energy Phys.} \textbf{2018}, {\emph{1810}}, 056.
  %%CITATION = doi:10.1007/JHEP10(2018)056;%%
  %14 citations counted in INSPIRE as of 10 Sep 2019  

 \bibitem{log2}
%\cite{Laddha:2018vbn}
%\bibitem{Laddha:2018vbn} 
 Laddha, A.; Sen, A. { Observational Signature of the Logarithmic Terms in the Soft Graviton Theorem}, 
  \emph{Phys.\ Rev.\ D} \textbf{2019}, {\emph{100}},  024009.
  %%CITATION = doi:10.1103/PhysRevD.100.024009;%%
  %10 citations counted in INSPIRE as of 04 Oct 2019

 \bibitem{log3}
%\cite{Sahoo:2018lxl}
%\bibitem{Sahoo:2018lxl} 
  Sahoo, B.; Sen, A. {Classical and Quantum Results on Logarithmic Terms in the Soft Theorem in Four Dimensions}. 
  \emph{J. High Energy Phys.} \textbf{2019}, {\emph{1902}}, 086.
  %%CITATION = doi:10.1007/JHEP02(2019)086;%%
  %17 citations counted in INSPIRE as of 10 Sep 2019
%
%%%%%%%%%%%%%%%%%%%%%%%%%%%%%%%
\bibitem{1911.05099}
Marotta, R.; Verma, M. {Soft Theorems from Compactification}, \emph{J. High Energy Phys.} \textbf{2020}, \emph{02}, 008.

\bibitem{2102.13180}
 Marotta, R.; Taronna, M.; Verma, M. {Revisiting higher-spin gyromagnetic couplings}, 
\emph{J. High Energy Phys.} \textbf{2021}, \emph{06}, 168.


%%%%%%%%%%%%%%%%%%%%%%%%%%%%%%%
\bibitem{BMBS1}
%\bibitem{BMS}
Bondi, H.; van Burg, M.G.J.; Metzner, A.W.K. %MDPI: there are not only one ref information in this ref, please separted into 2 refs and change the ref citation, please confirm and revise..
{Gravitational waves in general relativity. 7. Waves from axisymmetric isolated systems}. 
\emph{Proc.\ Roy.\ Soc.\ Lond.\ A} {\textbf{1962},  \emph{269}}, \emph{21}.
%%CITATION = PRSLA,A269,21;%%

\bibitem{BMBS2}
%
Sachs, R.K.
%``Gravitational waves in general relativity. 8. Waves in asymptotically flat space-times,''
\emph{Proc.\ Roy.\ Soc.\ Lond.\ A} {\textbf{1962},  \emph{270}}, 103.
%%CITATION = PRSLA,A270,103;%%
 
%%%%%%%%%%%%%%%%%%% 
\bibitem{Strominger}
Strominger, A. {Lectures on the Infrared Structure of Gravity and Gauge Theory}. \emph{arXiv} \textbf{2017},
arXiv:1703.05448  and references therein.
 

%%%%%%%%%%%%%%%%%%%%%%%%%%%
\bibitem{Ademollo:1975pf} 
 Ademollo, M.; D'Adda, A.; D'Auria, R.; Gliozzi, F.; Napolitano, E.; Sciuto, S.; Vecchia, P.D.
%\mbox{\href{http://dx.doi.org/10.1016/0550-3213(75)90491-5}{{\em 
{Soft Dilatons and Scale Renormalization in Dual Theories}.
\emph{Nucl.\ Phys.} \textbf{1975}, {\emph{B94}}, 221.
  %%CITATION = doi:10.1016/0550-3213(75)90491-5;%%
  %
  
\bibitem{Shapiro:1975cz}
 % \bibitem{Shapiro:1975cz}
Shapiro, J. {On the renormalization of Dual Models}.
%\href{http://dx.doi.org/ 10.1103/PhysRevD.11.2937}{
\emph{Phys. Rev.} {\textbf{1975}, \emph{ D11}}, 2937.
  %%CITATION = doi:10.1103/PhysRevD.11.2937;%%
     
     
 \bibitem{DMM1} 
   Vecchia, P.D.; Marotta, R.; Mojaza, M. {Soft theorem for the graviton, dilaton and the Kalb-Ramond field in the bosonic string}. 
   \emph{J. High Energy Phys.} \textbf{2015}, {\emph{1505}}, 137.
   %%CITATION = ARXIV:1502.05258;%%
   %8 citations counted in INSPIRE as of 13 Nov 2015
  %\cite{DiVecchia:2015jaq}
  
\bibitem{DMMN} 
  Vecchia, P.D.; Marotta, R.; Mojaza, M.; Nohle, J.
  { New soft theorems for the gravity dilaton and the Nambu-Goldstone dilaton at subsubleading order}. 
\emph{ Phys. Rev. D} {\textbf{2016},  \emph{93}}, 085015.
 
%\cite{DiVecchia:2016amo}
\bibitem{DMM2} 
  Vecchia, P.D.; Marotta, R.; Mojaza, M. { Subsubleading soft theorems of gravitons and dilatons in the bosonic string}.
  \emph{J. High Energy Phys.} \textbf{2016}, { \emph{1606}}, 054.
  %%CITATION = doi:10.1007/JHEP06(2016)054;%%
   %
   

   
\bibitem{1703.07885}Campiglia, M.; Coito, L.; Mizera, S. { Can scalars have asymptotic symmetries?} \emph{Phys. Rev. D} \textbf{2018}, \emph{97}, 46002.
%DOI:https://doi.org/10.1103/PhysRevD.97.046002   

\bibitem{1808.04845} Vecchia, P.D.; Marotta, R.; Mojaza, M. { Multiloop Soft Theorem for Gravitons and Dilatons in the Bosonic String}.  \emph{J. High Energy Phys.} \textbf{2019}, \emph{01}, 038. 
% https://doi.org/10.1007/JHEP01(2019)038
 
\bibitem{1907.01036}
 Vecchia, P.D.; Marotta, R.; Mojaza, M. {Multiloop soft theorem of the dilaton in the bosonic string}. \emph{Phys. Rev. D} \textbf{2019}, \emph{100}, 041902.
%DOI:https://doi.org/10.1103/PhysRevD.100.041902

\bibitem{geometric}
 Cheung, C.; Helset, A.; Parra-Martinez, J. {Geometric Soft Theorems}. \emph{arXiv} \textbf{2021}, arXiv:2111.03045.



%\cite{Adler:1964um}
\bibitem{Adler:1964um} 
  Adler, S.L. %MDPI: please check if ref 23 is same as ref 24, if they are same, please revise. They are two different references. 
 {Consistency conditions on the strong interactions implied by a partially conserved axial vector current}.
 \emph{ Phys.\ Rev.}  {\textbf{1965},  \emph{137}}, B1022.
  %%doi:10.1103/PhysRev.137.B1022
  %%CITATION = doi:10.1103/PhysRev.137.B1022;%%
  %411 citations counted in INSPIRE as of 28 Apr 2017


%\cite{Adler:1965ga}
\bibitem{Adler:1965ga} 
  Adler, S.L. {Consistency conditions on the strong interactions implied by a partially conserved axial-vector current. II}.
  \emph{Phys.\ Rev.}  {\textbf{1965},  \emph{139}}, B1638.
  %%doi:10.1103/PhysRev.139.B1638
  %%CITATION = doi:10.1103/PhysRev.139.B1638;%%
  


%\cite{Weinberg:1966gjf}
\bibitem{Weinberg:1966gjf} 
  Weinberg, S. {Current-Commutator Theory of Multiple Pion Production}.
  \emph{Phys.\ Rev.\ Lett.}  {\textbf{1966},  \emph{16}},  879.
  %%doi:10.1103/PhysRevLett.16.879
  %%CITATION = doi:10.1103/PhysRevLett.16.879;%%
  %94 citations counted in INSPIRE as of 28 Apr 2017

\bibitem{2009.07817}
Blas, D.; Camalich, J.M.; Oller, J.A. {Scalar resonance in graviton-graviton scattering at high-energies: The graviball}. \emph{Phys.  Lett. B} \textbf{2022}, \emph{827}, 136991.

\bibitem{0110285}
Low, I.; Manohar, A.V. %MDPI: ref 27 is same as ref 43,, please confirm and revise.
 {Spontaneously Broken Spacetime Symmetries and Goldstone's Theorem}. \emph{Phys. Rev. Lett.} \textbf{2002}, \emph{88}, 101602.
%%DOI: 10.1103/PhysRevLett.88.101602.

\bibitem{Callan:1970yg} 
  Callan, C.G., Jr.  {Broken scale invariance in scalar field theory}.
  \emph{Phys.\ Rev.\ D} \textbf{1970}, { \emph{2}}, 1541.
  %doi:10.1103/PhysRevD.2.1541
  %%CITATION = doi:10.1103/PhysRevD.2.1541;%%
  %980 citations counted in INSPIRE as of 28 Apr 2017


%\cite{Coleman:1970je}
\bibitem{Coleman:1970je} 
  Coleman, S.R.; Jackiw, R. {Why dilatation generators do not generate dilatations?}
  \emph{Ann. Phys.}  {\textbf{1971},  \emph{67}}, 552.
  %doi:10.1016/0003-4916(71)90153-9
  %%CITATION = doi:10.1016/0003-4916(71)90153-9;%%
  %311 citations counted in INSPIRE as of 28 Apr 2017


%\cite{Boels:2015pta}
\bibitem{Boels:2015pta} 
  Boels, R.H.; Wormsbecher, W. { Spontaneously broken conformal invariance in observables}.  \emph{arXiv} \textbf{2015},
  arXiv:1507.08162.
  %%CITATION = ARXIV:1507.08162;%%
  %11 citations counted in INSPIRE as of 28 Apr 2017

\bibitem{1705.06175}
 Vecchia, P.D.; Marotta, R.; Mojaza, M. {Double-soft behavior of the dilaton of spontaneously broken conformal invariance}.  \emph{J. High Energ. Phys.} \textbf{2017}, {\em 2017}, 1. 
  %%https://doi.org/10.1007/JHEP09(2017)001

%\cite{Bianchi:2016viy}
\bibitem{Bianchi:2016viy} 
  Bianchi, M.; Guerrieri, A.L.; Huang, Y.t.; Lee, C.J.; Wen, C. {Exploring soft constraints on effective actions}.
  \emph{J. High Energy Phys.} \textbf{2016}, {\em 1610}, 036.
  %%CITATION = doi:10.1007/JHEP10(2016)036;%%
  %7 citations counted in INSPIRE as of 28 Apr 2017  

  
  
%\cite{Dashen:1969ez}
\bibitem{Dashen:1969ez} 
  Dashen, R.F.; Weinstein, M. {Soft pions, chiral symmetry, and phenomenological lagrangians}.
  \emph{Phys.\ Rev.}  \textbf{1969}, {\em 183}, 1261.
%  doi:10.1103/PhysRev.183.1261
  %%CITATION = doi:10.1103/PhysRev.183.1261;%%
  %188 citations counted in INSPIRE as of 28 Apr 2017


%\cite{ArkaniHamed:2008gz}
\bibitem{ArkaniHamed:2008gz} 
  Arkani-Hamed, N.; Cachazo, F.; Kaplan, J. { What is the Simplest Quantum Field Theory?}
  \emph{J. High Energy Phys.} \textbf{2010}, {\em 1009}, 016.
  %%CITATION = doi:10.1007/JHEP09(2010)016;%%
  %359 citations counted in INSPIRE as of 28 Apr 2017


%\cite{Kampf:2013vha}
\bibitem{Kampf:2013vha} 
  Kampf, K.; Novotny, J.; Trnka, J. {Tree-level Amplitudes in the Nonlinear Sigma Model}.
  \emph{J. High Energy Phys.} \textbf{2013}, {\em 1305}, 032.
  %%CITATION = doi:10.1007/JHEP05(2013)032;%%
  %27 citations counted in INSPIRE as of 28 Apr 2017


%\cite{Low:2015ogb}
\bibitem{Low:2015ogb} 
  Low, I. { Double Soft Theorems and Shift Symmetry in Nonlinear Sigma Models},
\emph{  Phys.\ Rev.\ D} \textbf{2016}, {\em 93},  045032.
  %%CITATION = doi:10.1103/PhysRevD.93.045032;%%
  %12 citations counted in INSPIRE as of 28 Apr 2017


%\cite{Du:2015esa}
\bibitem{Du:2015esa} 
Du, Y.J.; Luo, H. { On single and double soft behaviors in NLSM}.   \emph{J. High Energy Phys.} \textbf{2015}, {\em 1508}, 058.
  %%CITATION = doi:10.1007/JHEP08(2015)058;%%
  %10 citations counted in INSPIRE as of 28 Apr 2017
  
\bibitem{DiVecchia:2017gfi}  
  Vecchia, P.D.; Marotta, R.; Mojaza, M. 
{The B-field soft theorem and its unification with the graviton and dilaton}. \emph{J. High Energy Phys.} \textbf{2017},  \emph{1710}, 017.
  
 \bibitem{1802.05999}	
Loebbert, F.; Mojaza, M.; Plefka, J. { Hidden Conformal Symmetry in Tree-Level Graviton Scattering}.
\emph{J. High Energy Phys.} \textbf{2018},  {\em 1805}, 208.
% 

\bibitem{2005.05877}
Marotta, R.; Mojaza, M. { Double-soft behavior of massless closed strings interacting with any number of closed string tachyons}. \emph{J. High Energy Phys.}  \textbf{2020}, \emph{08}, 083.

\bibitem{DiFrancesco}
Francesco, P.D.; Mathieu, P.; Senechal, D. {\em Conformal Field Theory}; Springer:  Berlin/Heidelberg, Germany, %newly added information, please confirm
1997. 
 https://doi.org/10.\linebreak1007/978-1-4612-2256-9.

\bibitem{Higashijima:1994}
Higashijima, K. { Nambu-Goldstone theorem for conformal symmetry}. In  {Proceedings of the XX International Colloquium on Group Theoretical Methods in Physics (GROUP 20), Toyonaka, Japan,  4--9 July} 1994;
 pp. 223-228.

%\bibitem{0110285}
%Low, I.; Manohar, A.V. { Spontaneously Broken Spacetime %Symmetries and Goldstone's Theorem}. \emph{Phys. Rev. Lett.} %\textbf{2002}, \emph{88}, 101602.



\bibitem{coleman} 
Coleman, S. {\em Aspects of Symmetry: Selected Erice Lectures}; Cambridge University Press: Cambridge, UK, 1988.

\bibitem{string1}
Polchinski, J. {\em String Theory. An Introduction to the Bosonic String};  Cambridge University Press: Cambridge, UK, 1998;  Volume 1.%MDPI: newly added information.


\bibitem{string2}
Polchinski, J. {\em String Theory. Superstring Theory and Beyond};  Cambridge University Press: Cambridge, UK, 1998;  %MDPI: newly added information.
Volume 2.


\bibitem{Hata:1992} 
Hata, H. {Softh Dilaton Theorem in String Field Theory}. \emph{Prog. Theor. Phys.} \textbf{1992},  \emph{88}, 1197.

\bibitem{Scherk:1971} Scherk, J. {Zero slope limit of the dual resonance model}. \emph{Nucl. Phys.} \textbf{1971}, {\emph{B31}}, 222.









\end{thebibliography}
\end{document}